\documentclass[10pt,letterpaper]{article}
\usepackage{opex3}
\usepackage{subfigure}
\usepackage{amssymb}
\usepackage{graphicx}
\usepackage{cite}

\begin{document}

\title{Symbiotic two-component gap solitons}
\author{Athikom Roeksabutr, $^{1}$ Thawatchai Mayteevarunyoo,$^{1\ast }$ and
Boris A. Malomed$^{2}$}
\address{
$^{1}$Department of Telecommunication Engineering,
Mahanakorn University of Technology, \\Bangkok 10530, Thailand\\
$^{2}$Department of Physical Electronics, School of Electrical
Engineering, \\Faculty of Engineering, Tel Aviv University, Tel Aviv
69978,
Israel\\
$^{*}$ \color{blue}\underline{thawatch@mut.ac.th}}

\begin{abstract}
We consider a two-component one-dimensional model of gap solitons (GSs),
which is based on two nonlinear Schr\"{o}dinger equations, coupled by
repulsive XPM (cross-phase-modulation) terms, in the absence of the SPM
(self-phase-modulation) nonlinearity. The equations include a periodic
potential acting on both components, thus giving rise to GSs of the
``symbiotic" type, which exist solely due to the repulsive interaction
between the two components. The model may be implemented for ``holographic
solitons" in optics, and in binary bosonic or fermionic gases trapped in the
optical lattice. Fundamental symbiotic GSs are constructed, and their
stability is investigated, in the first two finite bandgaps of the
underlying spectrum. Symmetric solitons are destabilized, including their
entire family in the second bandgap, by symmetry-breaking perturbations
above a critical value of the total power. Asymmetric solitons of intra-gap
and inter-gap types are studied too, with the propagation constants of the
two components falling into the same or different bandgaps, respectively.
The increase of the asymmetry between the components leads to shrinkage of
the stability areas of the GSs. Inter-gap GSs are stable only in a strongly
asymmetric form, in which the first-bandgap component is a dominating one.
Intra-gap solitons are unstable in the second bandgap. Unstable
two-component GSs are transformed into persistent breathers. In addition to
systematic numerical considerations, analytical results are obtained by
means of an extended (``tailed") Thomas-Fermi approximation (TFA).
\end{abstract}








\section{Introduction}

Studies of solitons in spatially periodic (lattice) potentials have grown
into a vast area of research, with profoundly important applications to
nonlinear optics, plasmonics, and matter waves in quantum gases, as outlined
in recent reviews \cite{Rev0,Rev1,Rev2,Rev3}. In ultracold bosonic and
fermionic gases, periodic potentials can be created, in the form of optical
lattices, by coherent laser beams illuminating the gas in opposite
directions \cite{Pit,Stoof,Morsch}. Effective lattice potentials for optical
waves are induced by photonic crystals, which are built as permanent
structures by means of various techniques \cite{Rev1,PhotCryst,Jena}, or as
laser-induced virtual structures in photorefractive crystals \cite{Moti}.
Parallel to the progress in the experiments, the study of the interplay
between the nonlinearity and periodic potentials has been an incentive for
the rapid developments of theoretical methods \cite{Yang,Pelinovsky}. Both
the experimental and theoretical results reveal that solitons can be created
in lattice potentials, if they do not exist in the uniform space [this is
the case of gap solitons (GSs) supported by the self-defocusing
nonlinearity, see original works \cite{GS-Mario,GS-Kivshar,HS,radial} and
reviews \cite{Brazhnyy,Morsch}], and solitons may be stabilized, if they are
unstable without the lattice (multidimensional solitons in the case of
self-focusing, as shown in Refs. \cite%
{BBB,YM,YM2,BBB2,Barcelona,BBB3,BBB4,Thawatchai}, see also reviews \cite%
{Rev0,Rev2,Rev3}). The stability of GSs has been studied in detail
too---chiefly, close to edges of the corresponding bandgaps---in one \cite%
{stability1,stability2,JC} and two \cite{stability2D} dimensions alike.

An essential extension of the theme is the study of two-component solitons
in lattice potentials. In particular, if both the self-phase-modulation and
cross-phase-modulation (SPM and XPM) nonlinearities, i.e., intra- and
inter-species interactions, are repulsive, one can construct two-component
GSs of \textit{intra-gap} and \textit{inter-gap} types, with chemical
potentials of the components (or propagation constants, in terms of optical
media) falling, respectively, into the same or different bandgaps of the
underlying linear spectrum \cite{Arik,Sadhan}. In the case of the attractive
SPM, a family of stable \textit{semi-gap solitons} was found too, with one
component residing in the infinite gap, while the other stays in a finite
bandgap \cite{Sadhan}. The GSs supported by the XPM repulsion dominating
over the intrinsic (SPM-mediated) attraction may be regarded as an example
of \textit{symbiotic solitons}. In the free space (without the lattice
potential), symbiotic solitons are supported by the XPM attraction between
their two components, despite the action of the repulsive SPM in each one
\cite{symbio1,symbio2,symbio3}. This mechanism may be additionally enhanced
by the linear coupling (interconversion) between the components \cite%
{Sadhan2}. Another case of the ``symbiosis" was reported in Ref. \cite{Olga}%
, where the action of the lattice potential on a single component was
sufficient for the stabilization of two-dimensional (2D) two-component
solitons against the collapse, the stabilizing effect of the lattice on the
second component being mediated by the XPM interaction. In addition, the
attraction between the components, competing with the intrinsic repulsion,
may cause spatial splitting between two components of the GS, as for these
components, whose effective masses are negative \cite{HS}, the attractive
interaction potential gives rise to a repulsion force \cite{Michal,Sadhan}.

The ultimate form of the model which gives rise to two-component GSs of the
symbiotic type is the one with no intra-species nonlinearity, the formation
of the GSs being accounted for by the interplay of the repulsion between the
components and the lattice potential acting on both of them. In optics, the
setting with the XPM-only interactions is known in the form of the
``holographic nonlinearity", which can be induced in photorefractive
crystals for a pair of coherent beams with a small angle between their wave
vectors, giving rise to single- \cite{holo,holo2} and double-peak \cite%
{holoKiv} solitons. Both beams are made by splitting a single laser signal,
hence the power ratio between them (which is essential for the analysis
reported below) can be varied by changing the splitting conditions. The
creation of 2D spatial ``holographic solitons" in a
photorefractive-photovoltaic crystal with the self-focusing nonlinearity was
demonstrated in Ref. \cite{holo3} (such solitons are stable, as the collapse
is arrested by the saturation of the self-focusing). To implement the
situation considered here, the sign of the nonlinearity may be switched to
self-defocusing by the reversal of the bias voltage, and the effective
lattice potential may be induced by implanting appropriate dopants, with the
concentration periodically modulated in one direction, which will render the
setting quasi-one-dimensional.

In binary bosonic gases, a similar setting may be realized by switching off
the SPM nonlinearity with the help of the Feshbach resonance, although one
may need to apply two different spatially uniform control fields (one
magnetic and one optical) to do it simultaneously in both components. On the
other hand, the same setting is natural for a mixture of two fermionic
components with the repulsive interaction between them, which may represent
two states of the same atomic species, with different values of the total
atomic spin ($F$). If spins of both components are polarized by an external
magnetic field, the SPM\ nonlinearity will be completely suppressed by the
Pauli blockade while the inter-component interaction remains active \cite%
{Stoof}, hence the setting may be described by a pair of Schr\"{o}dinger
equations for the two wave functions, coupled by XPM terms.

The objective of this work is to present basic families of one-dimensional
symbiotic GSs, supported solely by the repulsive XPM nonlinearity in the
combination with the lattice potential, and analyze their stability, via the
computation of eigenvalues for small perturbations and direct simulations of
the perturbed evolution. The difference from the previously studied models
of symbiotic solitons \cite{Arik,Sadhan} is that the solitons where created
there by the SPM nonlinearity separately in each component, while the XPM
interaction determined the interaction between them and a possibility of
creating two-component bound states. Here, the two-component GSs may exist
solely due to the repulsive XPM interactions between the components.

We conclude that the symmetric solitons, built of equal components, are
destabilized by symmetry-breaking perturbations above a certain critical
value of the soliton's power. The analysis is chiefly focused on asymmetric
symbiotic GSs, and on breathers into which unstable solitons are
transformed. The model is introduced in Section II, which is followed by the
analytical approximation presented in Section III. It is an extended
(\textquotedblleft tailed") version of the Thomas-Fermi approximation, TFA,
which may be applied to other models too. In Section IV, we report
systematic numerical results obtained for fundamental solitons of both the
intra- and inter-gap types, hosted by the first two finite bandgaps of the
system's spectrum. The most essential findings are summarized in the form of
plots showing the change of the GS stability region with the variation of
the degree of asymmetry of the two-component symbiotic solitons, which is a
new feature exhibited by the present system. In particular, the stability
area of intra-gap solitons shrinks with the increase of the asymmetry, while
inter-gap solitons may be stable only if the asymmetry is large enough, in
favor of the first-bandgap component, and intra-gap solitons in the second
bandgap are completely unstable. The paper is concluded by Section V.

\section{The model}

The model outlined above is represented by the system of XPM-coupled Schr%
\"{o}dinger equations for local amplitudes of co-propagating electromagnetic
waves in the planar optical waveguide, $u(x,z)$ and $v(x,z)$, where $x$ and $%
z$ are the transverse coordinate and propagation distance, without the SPM
terms, and with the lattice potential of depth $2\varepsilon >0$ acting on
both components:%
\begin{eqnarray}
i\frac{\partial u}{\partial z}+\frac{1}{2}\frac{\partial ^{2}u}{\partial
x^{2}}-|v|^{2}u+\varepsilon \cos (2x)u &=&0,  \label{1Dpsi} \\
i\frac{\partial v}{\partial z}+\frac{1}{2}\frac{\partial ^{2}v}{\partial
x^{2}}-|u|^{2}v+\varepsilon \cos (2x)v &=&0.  \label{1Dphi}
\end{eqnarray}%
The variables are scaled so as to make the lattice period equal to $\pi $,
and the coefficients in front of the diffraction and XPM terms equal to $1$.
In the case of matter waves, $u$ and $v$ are wave functions of the two
components, and $z$ is replaced by time $t$. Direct simulations of Eqs. (\ref%
{1Dpsi}) and (\ref{1Dphi}) were performed with the help of the split-step
Fourier-transform technique.
\begin{figure}[tbp]
\centering\includegraphics[width=3in]{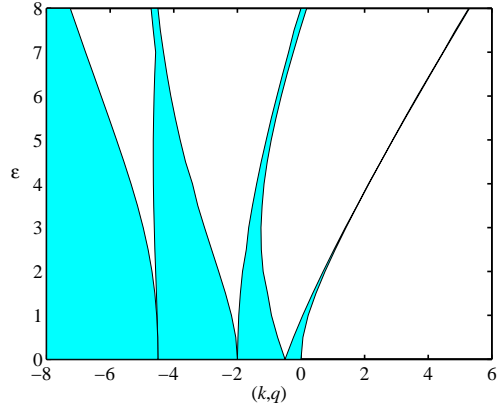}
\caption{The identical bandgap structures produced by the linearization of
Eqs. (\protect\ref{ODE1}) and (\protect\ref{ODE2}) for $\protect\varepsilon %
=6$. Shaded areas are occupied by the Bloch bands, where gap solitons do not
exist.}
\label{fig1}
\end{figure}

Stationary solutions to Eqs. (\ref{1Dpsi}), (\ref{1Dphi}) are looked as%
\begin{equation}
u\left( x,z\right) =e^{ikz}U(x),~v\left( x,z\right) =e^{iqz}V(x),  \label{GS}
\end{equation}%
where the real propagation constants, $k$ and $q$, are different, in the
general case, and real functions $U(x)$ and $V(x)$ obey equations%
\begin{eqnarray}
-kU+\frac{1}{2}U^{\prime \prime }-V^{2}U+\varepsilon \cos (2x)U &=&0,
\label{ODE1} \\
-qV+\frac{1}{2}V^{\prime \prime }-U^{2}V+\varepsilon \cos (2x)V &=&0,
\label{ODE2}
\end{eqnarray}%
with the prime standing for $d/dx$. Numerical solutions to Eqs. (\ref{ODE1})
and (\ref{ODE2}) were obtained by means of the Newton's method.

Solitons are characterized by the total power,%
\begin{equation}
P=\int_{-\infty }^{+\infty }(\left\vert U\right\vert ^{2}+\left\vert
V\right\vert ^{2})dx\equiv P_{u}+P_{v},  \label{P}
\end{equation}%
with both $P_{u}$ and $P_{v}$ being dynamical invariants of Eqs. (\ref{1Dpsi}%
), (\ref{1Dphi}), and by the \textit{asymmetry ratio},%
\begin{equation}
R=\left( P_{u}-P_{v}\right) /\left( P_{u}+P_{v}\right) .  \label{R}
\end{equation}%
The total power and asymmetry may be naturally considered as functions of
the propagation constants, $k$ and $q$.

The well-known bandgap spectrum of the linearized version of Eqs. (\ref{ODE1}%
), (\ref{ODE2}) (see, e.g., book \cite{Yang}) is displayed in Fig. \ref{fig1}%
, the right edge of the first finite bandgap being
\begin{equation}
k_{\max }\left( \varepsilon =6\right) \approx 3.75.  \label{k_max}
\end{equation}%
The location of GSs is identified with respect to bandgaps of the spectrum.
In this work, results are reported for composite GSs whose two components
belong to the first and second finite bandgaps.

Stability of the stationary solutions can be investigated by means of the
linearization against small perturbations \cite%
{stability,stability1,stability2}. To this end, perturbed solutions of Eqs. (%
\ref{1Dpsi}) and (\ref{1Dphi}) are looked for as%
\begin{eqnarray}
u\left( x,z\right) &=&e^{ikz}\left[ U(x)+u_{1}(x)e^{-i\lambda z}+u_{2}^{\ast
}(x)e^{i\lambda ^{\ast }z}\right] ,  \nonumber \\
v\left( x,z\right) &=&e^{iqz}\left[ V(x)+v_{1}(x)e^{-i\lambda z}+v_{2}^{\ast
}(x)e^{i\lambda ^{\ast }z}\right] ,  \label{pert}
\end{eqnarray}%
where $u_{1,2}$ and $v_{1,2}$ are wave functions of infinitesimal
perturbations, and $\lambda $ is the respective instability growth rate,
which may be complex (the asterisk stands for the complex conjugate). The
instability takes place if there is at least one eigenvalue with \textrm{Im}$%
\left( \lambda \right) >0$. The substitution of ansatz Eqs. (\ref{pert})
into Eqs. (\ref{1Dpsi}), (\ref{1Dphi}) and the linearization with respect to
the small perturbations leads to the eigenvalue problem based on the
following equations:%
\begin{eqnarray}
qv_{1}-\frac{1}{2}v_{1}^{\prime \prime }+U^{2}(x)v_{1}+U(x)V(x)\left(
u_{1}+u_{2}\right) -\varepsilon \cos (2x)v_{1} &=&\lambda v_{1},  \label{u1}
\\
-qv_{2}+\frac{1}{2}v_{2}^{\prime \prime }-U^{2}(x)v_{2}-U(x)V(x)\left(
u_{1}+u_{2}\right) +\varepsilon \cos (2x)v_{2} &=&\lambda v_{2},  \label{u2}
\\
ku_{1}-\frac{1}{2}u_{1}^{\prime \prime }+V^{2}(x)u_{1}+U(x)V(x)\left(
v_{1}+v_{2}\right) -\varepsilon \cos (2x)u_{1} &=&\lambda u_{1},  \label{v1}
\\
-ku_{2}+\frac{1}{2}u_{2}^{\prime \prime }-V^{2}(x)u_{2}-U(x)V(x)\left(
v_{1}+v_{2}\right) +\varepsilon \cos (2x)u_{2} &=&\lambda u_{2}.  \label{v2}
\end{eqnarray}%
These equations were solved by means of the fourth-order center-difference
numerical scheme.

Results for the shape and stability of GSs of different types are presented
below for lattice strength $\varepsilon =6$, which adequately represents the
generic case. Note, in particular, that Fig. \ref{fig1} was plotted for this
value of the lattice-potential's strength.

\section{The extended Thomas-Fermi approximation}

It is well known that, close to edges of the bandgap, GSs feature an
undulating shape, which may be approximated by a Bloch wave function
modulated by a slowly varying envelope \cite{GS-Mario,HS}. On the other
hand, deeper inside the bandgap, the GSs are strongly localized (see, e.g.,
Figs. \ref{fig4} and \ref{fig7} below), which suggests to approximate them
by means of the variational method based on the Gaussian ansatz \cite%
{Sadhan,PhysicaD}. This approximation was quite efficient for the
description of GSs in single-component models \cite{PhysicaD}, while for
two-component systems it becomes cumbersome \cite{Arik,Sadhan}. 
%

Explicit analytical results for well-localized patterns can be obtained by
means of the TFA \cite{Pit}, which, in the simplest case, neglects the
kinetic-energy terms, $U^{\prime \prime }$ and $V^{\prime \prime }$, in Eqs.
(\ref{ODE1}), (\ref{ODE2}). Assuming, for the sake of the definiteness, $q<k$
and also $|k|<\varepsilon $ (the TFA is irrelevant for $|k|>\varepsilon $),
the approximation yields the fields inside the \textit{inner layer} of the
solution:%
\begin{equation}
\left\{
\begin{array}{c}
U^{2}(x) \\
V^{2}(x)%
\end{array}%
\right\} _{\mathrm{inner}}=\left\{
\begin{array}{c}
\varepsilon \cos \left( 2x\right) -q \\
\varepsilon \cos \left( 2x\right) -k%
\end{array}%
\right\} ,~\mathrm{at}~~|x|<x_{0}\equiv \frac{1}{2}\cos ^{-1}\left( \frac{k}{%
\varepsilon }\right) .  \label{in}
\end{equation}%
Thus, the TFA predicts the core part of the solution in the form of peaks in
the two components with the same width, $2x_{0}$, but different heights, $%
\left\{ U^{2},V^{2}\right\} _{\max }=\left\{ \varepsilon -q,\varepsilon
-k\right\} $. This structure complies with numerically generated examples of
asymmetric solitons displayed below in Figs. \ref{fig7}(a,b).
\begin{figure}[tbp]
\centering\subfigure[]{\includegraphics[width=2.5in]{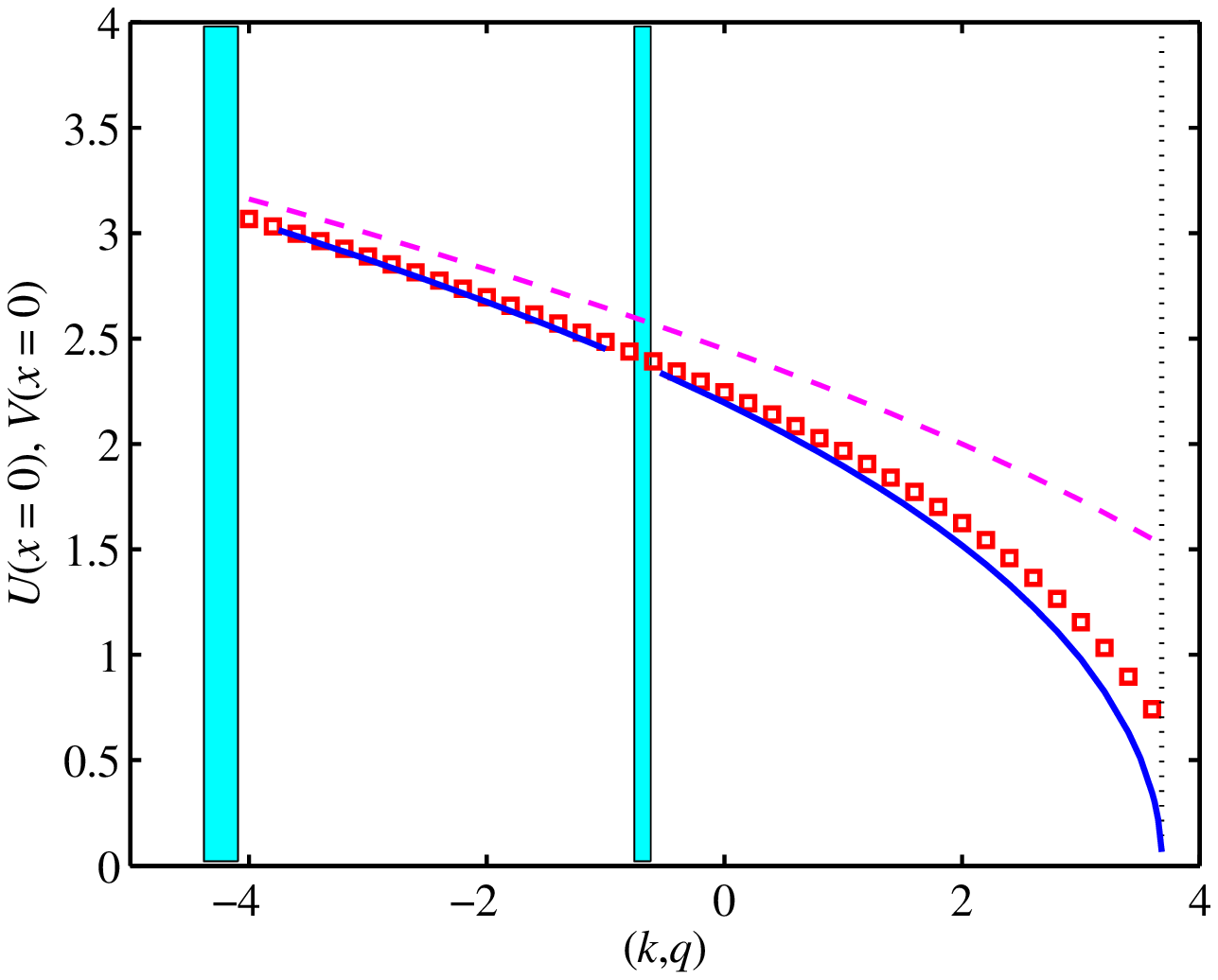}}%
\subfigure[]{\includegraphics[width=2.5in]{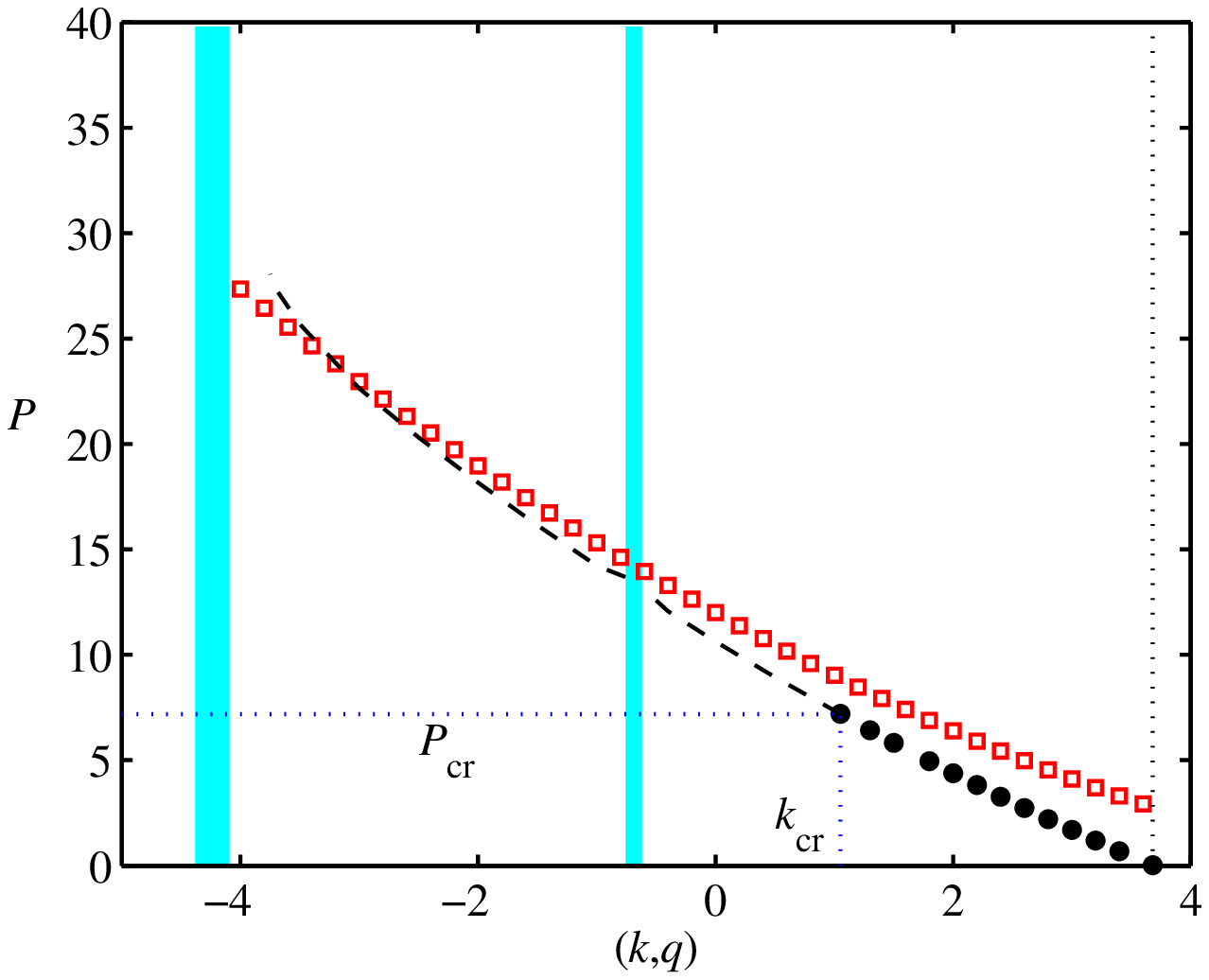}}
\caption{(a) The continuous (blue) curves show the numerically found
amplitude of the fundamental symmetric gap solitons (with equal components),
versus propagation constant $k=q$, in the first and second bandgaps at $%
\protect\varepsilon =6.0$. The chain of symbols is the analytical
approximation for the same dependence, as produced by the improved TFA in
the form of Eq. (\protect\ref{corr}). The dashed curve is the result of the
usual TFA, which corresponds to Eq. (\protect\ref{corr}) without the
correction (second) terms. (b) Total power $P$ for the same soliton
families, whose stable and unstable portions are designated by the bold
dotted and dashed lines, respectively. The latter one is destabilized by
symmetry-breaking perturbations, while the entire family is stable in the
framework of the single-component model. Coordinates of the
stability/instability border are given by Eq. (\protect\ref{cr}). The chain
of squares shows the analytical dependence produced by the TFA, see Eq. (%
\protect\ref{PTF}).}
\label{fig2}
\end{figure}

Further, the expansion of expressions (\ref{in}) around the soliton's center
($x=0$), yields%
\begin{equation}
\left\{
\begin{array}{c}
U(x) \\
V(x)%
\end{array}%
\right\} \approx \left\{
\begin{array}{c}
\sqrt{\varepsilon -q}-\left( \varepsilon /\sqrt{\varepsilon -q}\right) x^{2},
\\
\sqrt{\varepsilon -k}-\left( \varepsilon /\sqrt{\varepsilon -k}\right) x^{2}.%
\end{array}%
\right\} ~  \label{expand}
\end{equation}%
The substitution of the second derivatives of the fields at $x=0$,
calculated as per Eq. (\ref{expand}), into Eqs. (\ref{ODE1}), (\ref{ODE2})
yields a corrected expression for the soliton's amplitudes:%
\begin{equation}
\left\{
\begin{array}{c}
U(x=0) \\
V(x=0)%
\end{array}%
\right\} \approx \left\{
\begin{array}{c}
\sqrt{\varepsilon -q}-\varepsilon \left[ 2\left( \varepsilon -k\right) \sqrt{%
\varepsilon -q}\right] ^{-1}, \\
\sqrt{\varepsilon -k}-\varepsilon \left[ 2\left( \varepsilon -q\right) \sqrt{%
\varepsilon -k}\right] ^{-1},%
\end{array}%
\right\} ~  \label{corr}
\end{equation}%
along with the condition for the applicability of the TFA:%
\begin{equation}
\varepsilon \ll \left( \varepsilon -k\right) \left( \varepsilon -q\right) .~
\label{<<}
\end{equation}%
For the symmetric-GS\ families in the two first finite bandgaps, the
amplitude predicted by the improved TFA in the form of Eq. (\ref{corr}) is
displayed, as a function of $k=q$, in Fig. \ref{fig2}(a) and compared to its
numerically found counterpart. It is worthy to note that the correction
terms in Eq. (\ref{corr}) essentially improve the agreement of the TFA
prediction with the numerical findings.

At $|x|>x_{0}$, the TFA gives $V(x)=0$, which is a continuous extension of
the respective expression Eq. (\ref{in}) in the $V$ component, while the
continuity of fields $U(x)$ and $U^{\prime }(x)$ makes it necessary to match
the respective expression Eq. (\ref{in}) to \textquotedblleft tails", which,
in the lowest approximation, satisfy equation $U^{\prime \prime }=0$. The
continuity is provided by the following tail solution:%
\begin{equation}
\left\{ U^{2}(x)\right\} _{\mathrm{outer}}=\left\{
\begin{array}{c}
\left[ \sqrt{k-q}-\sqrt{\left( \varepsilon ^{2}-k^{2}\right) /\left(
k-q\right) }\left( |x|-x_{0}\right) \right] ^{2},~ \\
\mathrm{at}~~0<|x|-x_{0}<\left( k-q\right) /\sqrt{\varepsilon ^{2}-k^{2}};
\\
0,~\mathrm{at}~~|x|-x_{0}>\left( k-q\right) /\sqrt{\varepsilon ^{2}-k^{2}}.%
\end{array}%
\right\}  \label{outer}
\end{equation}

The integration of expressions Eq. (\ref{in}) and Eq. (\ref{outer}) yields
the following approximation for the powers of the two components:%
\begin{equation}
\left\{
\begin{array}{c}
P_{u} \\
P_{v}%
\end{array}%
\right\} _{\mathrm{TFA}}=\left\{
\begin{array}{c}
\sqrt{\varepsilon ^{2}-k^{2}}-q\cos ^{-1}\left( k/\varepsilon \right) \\
+(2/3)\left( k-q\right) ^{2}/\sqrt{\varepsilon ^{2}-k^{2}}; \\
\sqrt{\varepsilon ^{2}-k^{2}}-k\cos ^{-1}\left( k/\varepsilon \right) .%
\end{array}%
\right\}  \label{PP}
\end{equation}%
The substitution of approximation Eqs. (\ref{PP}) into definitions Eq. (\ref%
{P}) and Eq. (\ref{R}) of the total power and asymmetry demonstrates an
agreement with numerical results. For instance, the slope of the curve $R(q)$
for the intra-gap GSs at fixed $k$ (see Fig. \ref{fig8} below) at the
symmetry point ($k=q$), as predicted by Eqs. (\ref{PP}) for $\varepsilon =6$
and $k=1$, is $\left( \partial R/\partial q\right) |_{q=k}\approx -0.155,$
while its numerically found counterpart is $\approx -0.160$. Further, the
analysis of Eqs. (\ref{PP}) readily demonstrates that the strongly
asymmetric solitons may exist up to the limit of $R\rightarrow 1$, which is
corroborated by the existence area for the intra-gap solitons shown below in
Fig. \ref{fig11}(b).
\begin{figure}[tbp]
\centering\includegraphics[width=3in]{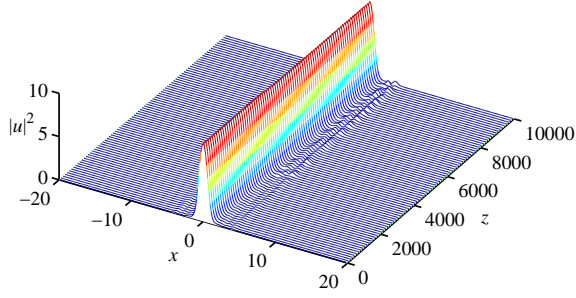}
\caption{The evolution of a weakly unstable single-component fundamental
soliton at $k=-3.7$.}
\label{fig3}
\end{figure}

Another corollary of Eqs. (\ref{PP}) is the prediction for the total power
for the symmetric solitons,%
\begin{equation}
P(k=q)=2\left[ \sqrt{\varepsilon ^{2}-k^{2}}-k\cos ^{-1}\left( k/\varepsilon
\right) \right] ,  \label{PTF}
\end{equation}%
which is plotted in Fig. \ref{fig2}(b), along with its numerically found
counterpart. Although the TFA does not predict edges of the bandgaps, the
overall analytical prediction for $P(k)$ runs quite close to the numerical
curve, except for near the right edge, where, indeed condition Eq. (\ref{<<}%
) does not hold for $\varepsilon =6$ and $k=3.75$, see Eq. (\ref{k_max}).
Note that this very simple analytical approximation was not derived before
in numerous works dealing with single-component GSs.
\begin{figure}[tbp]
\centering\includegraphics[width=3in]{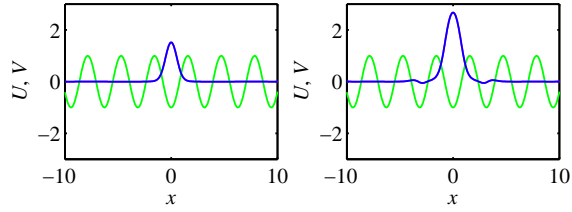}
\caption{Examples of fundamental symmetric gap solitons found in the first
and second bandgaps, for $k=q=2.0$ and $k=q=-2.0$ (left and right panels,
respectively). Here and in similar figures below, the background pattern
(green sinusoid) represents the underlying periodic potential. Both solitons
are stable as solutions of the single-component model, but only the one
corresponding to $k=q=2.0$ remains stable in the two-component system, while
its counterpart pertaining to $k=q=-2$ is destabilized by symmetry-breaking
perturbations.}
\label{fig4}
\end{figure}
\begin{figure}[tbp]
\centering\subfigure[]{\includegraphics[width=2.5in]{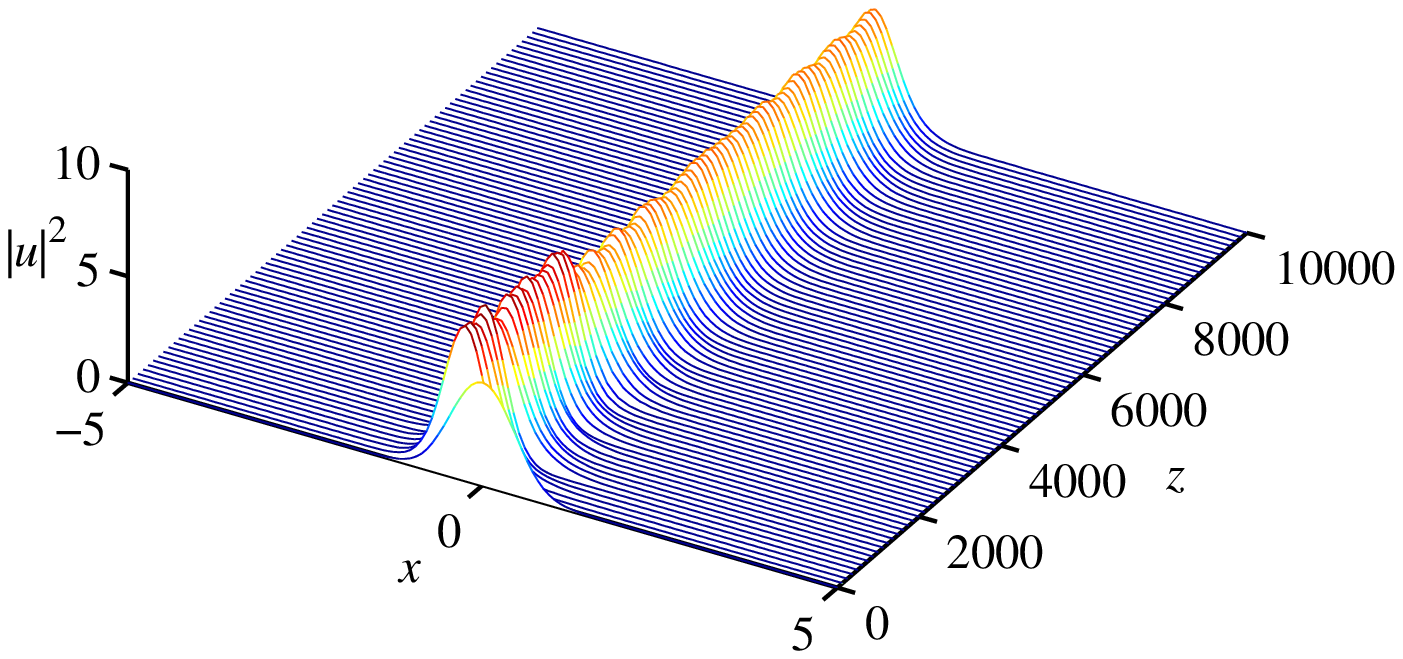}}%
\subfigure[]{\includegraphics[width=2.5in]{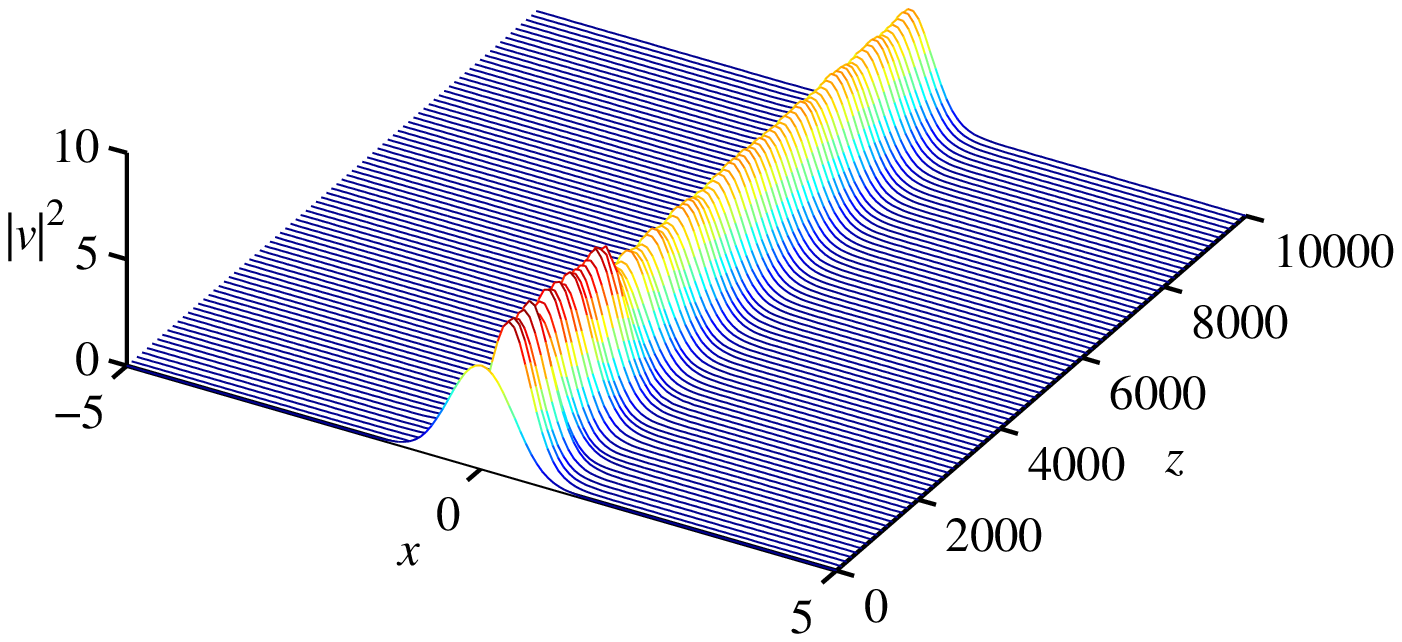}} \subfigure[]{%
\includegraphics[width=2.5in]{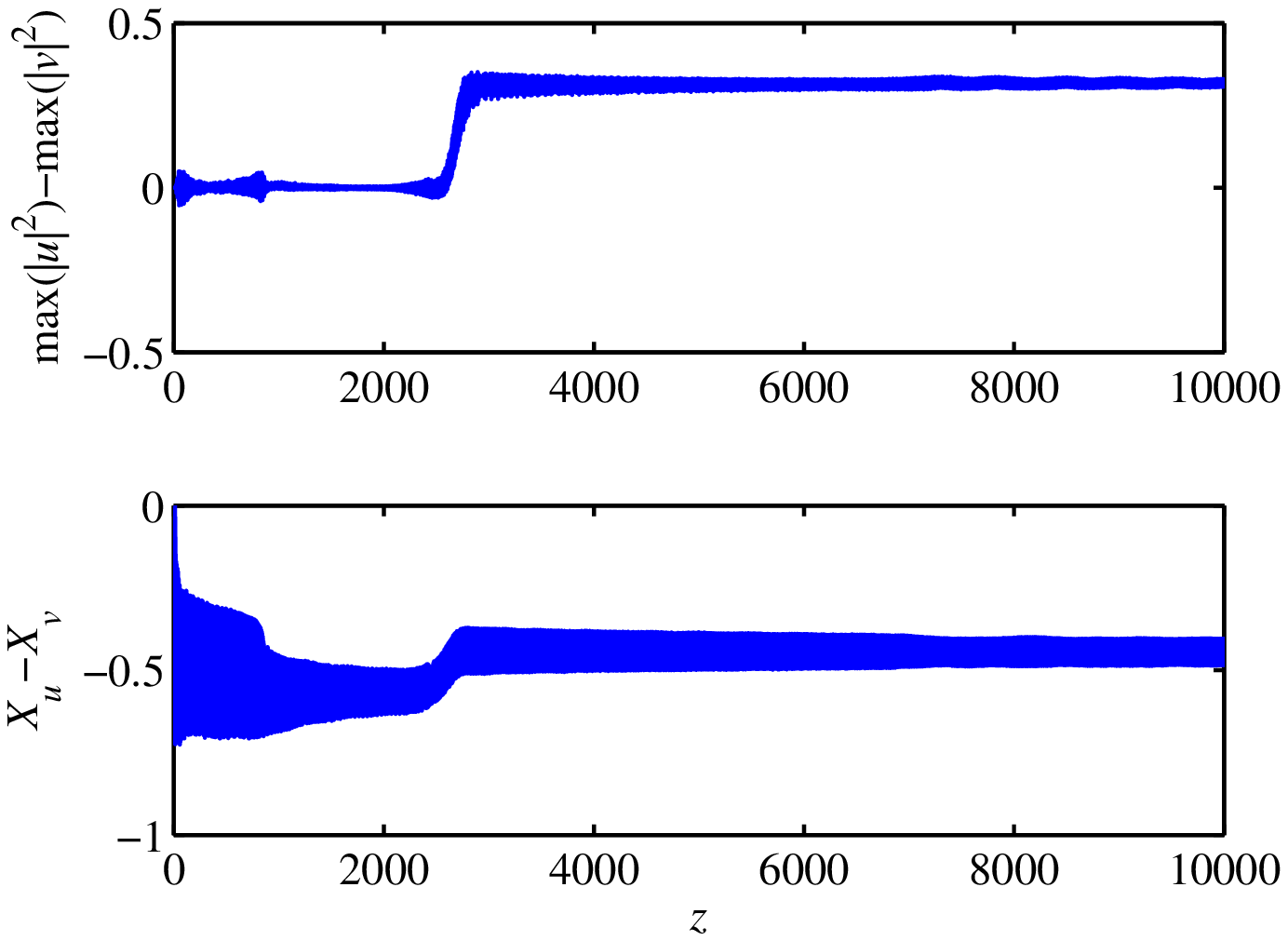}}
\caption{(a) and (b) The spontaneous transformation of an unstable symmetric
fundamental soliton for $u$- and $v$-component, in the first bandgap, with $%
k=q=0$, into a stable asymmetric breather. (c) The top and bottom plots
display, respectively, the evolution of the peak-power difference, $\max
(\left\vert u\left( x,z\right) \right\vert ^{2})-\max (\left\vert u\left(
x,z\right) \right\vert ^{2})$, and the separation between centers of the two
components, $X_{u}$ and $X_{v}$, which as per Eq. (\protect\ref{X}).}
\label{fig5}
\end{figure}
\begin{figure}[tbp]
\centering\subfigure[]{\includegraphics[width=2.5in]{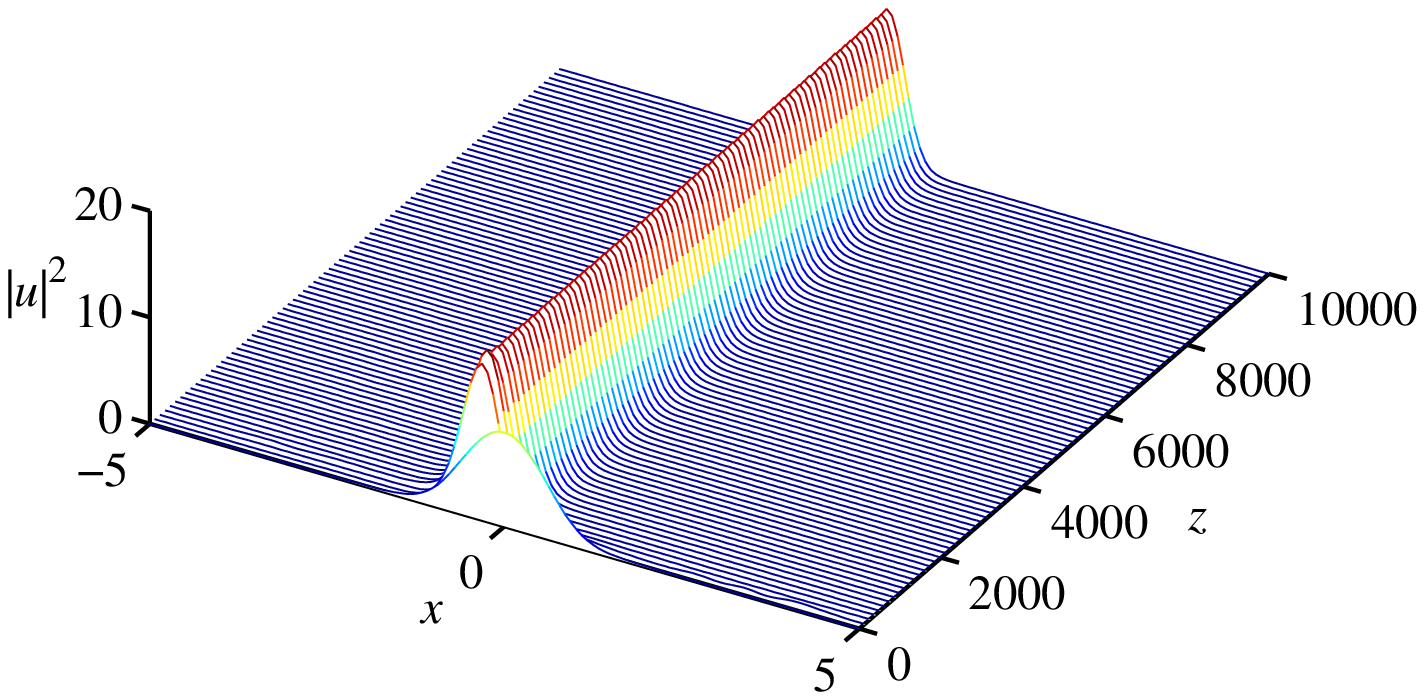}}%
\subfigure[]{\includegraphics[width=2.5in]{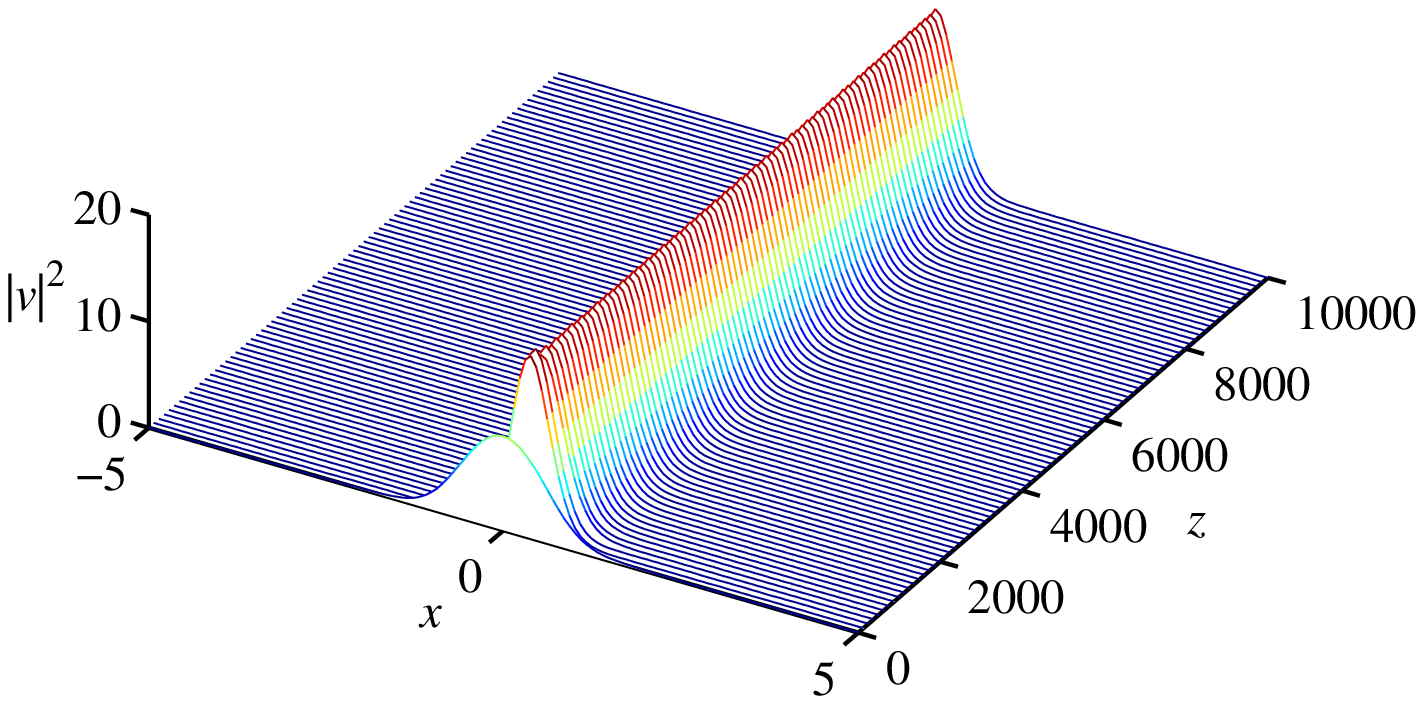}}
\caption{(a,b) The same as in Fig. \protect\ref{fig5}(a,b), but for an
unstable soliton in the second bandgap, with $k=q=-3.5$.}
\label{fig6}
\end{figure}
\begin{figure}[tbp]
\centering\subfigure[]{\includegraphics[width=2in]{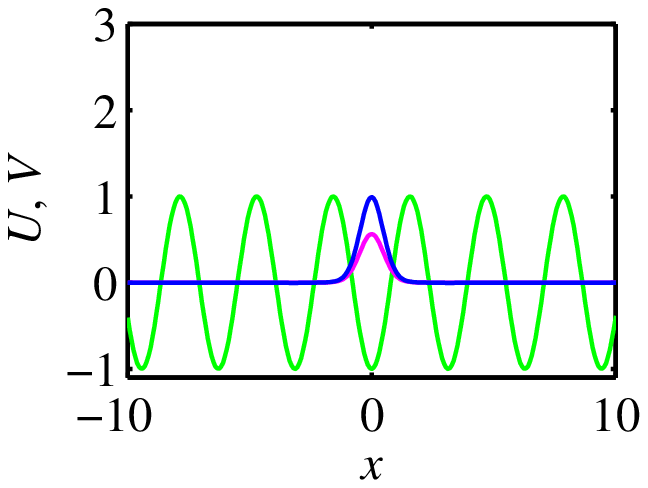}}%
\subfigure[]{\includegraphics[width=2in]{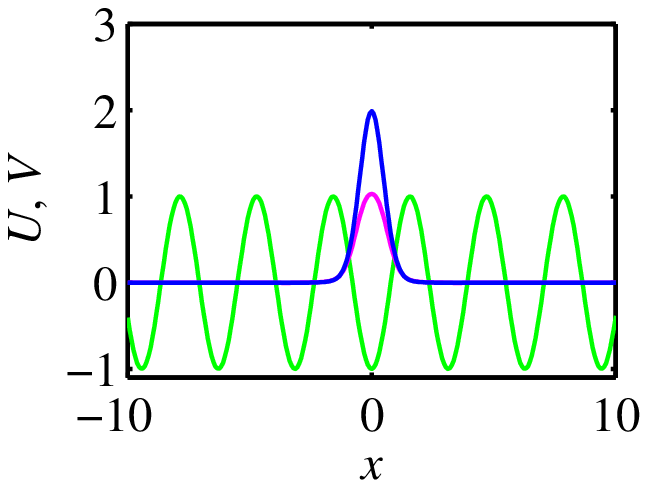}} \subfigure[]{%
\includegraphics[width=2in]{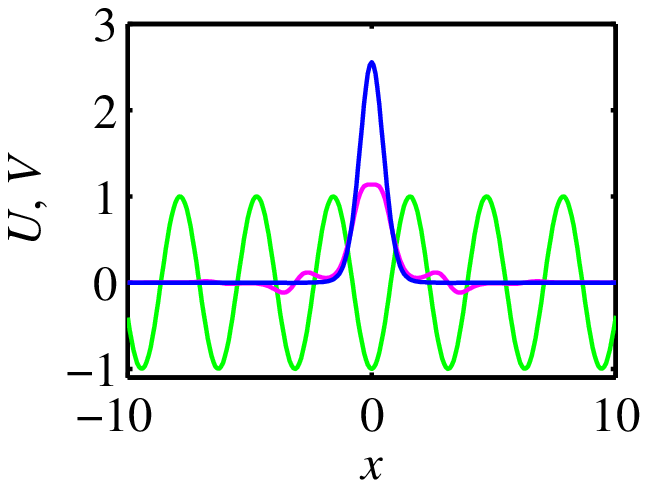}}
\caption{Examples of stable solitons of the intra-gap type found in the
first finite bandgap, with fixed asymmetry $R=-0.5$: (a) $k=3$ and $q=3.4601$%
; (b) $k=1$ and $q=2.843$; (c) $k=-0.5$ and $q=2.5116$. Fields $U(x)$ and $%
V(x)$, which pertain to propagation constants $k$ and $q$, are shown,
respectively, by the magenta (lower) and blue (higher) profiles.}
\label{fig7}
\end{figure}

\section{Results of the numerical analysis}

\subsection{Symmetric solitons}

Obviously, the shape of symmetric solitons, built of two equal components
[with $k=q$ and $U(x)=V(x)$, see Eq. (\ref{GS})], is identical to that of
GSs in the single-component model. However, there is a drastic difference in
the stability of the symmetric GSs between the single- and two-component
systems. Almost the entire symmetric family is stable against symmetric
perturbations, i.e., it is stable in the framework of the single-component
equation (in agreement with previously known results \cite{Yang}), except
for a weak oscillatory instability, accounted for by quartets of
complex-conjugate eigenvalues, in the form of $\lambda =\pm i\mathrm{Im}%
\left( \lambda \right) \pm \mathrm{Re}\left( \lambda \right) $ (with two
mutually independent signs $\pm $), which appears near the left edge of the
second bandgap---namely, at $k<k_{\min }\approx -3.45$. An example the
development of the latter instability is displayed below in Fig. \ref{fig3}.

On the other hand, Fig. \ref{fig2} demonstrates that a considerable part of
the family in the first finite bandgap, and the entire family in the second
bandgap are unstable against symmetry-breaking perturbations in the
two-component system. The boundary separating the stable and unstable
subfamilies of the fundamental symmetric GSs in the first finite bandgap
corresponds to the power and propagation constants is found at
\begin{equation}
P_{\mathrm{cr}}\approx 7.19,~k_{\mathrm{cr}}\approx 1.05,  \label{cr}
\end{equation}%
the symmetric solitons being stable in the intervals of $0<P<7.19$, $%
1.05<k<k_{\max }\approx 3.75$ [see Eq. (\ref{k_max})]. These results were
produced by a numerical solution of Eqs. (\ref{u1})-(\ref{v2}) (the
instability is oscillatory, characterized by complex eigenvalues).

Typical examples of stable and unstable fundamental symmetric GSs, found in
the first and second bandgaps (not too close to their edges), are displayed
in Fig. \ref{fig4}. Further, direct simulations demonstrate that the
evolution transforms the unstable symmetric solitons into persistent
localized breathers, as shown in Figs. \ref{fig5} and \ref{fig6}, in
accordance with the fact that the corresponding instability eigenvalues are
complex. Although the emerging breather keeps the value of $R=0$, see Eq. (%
\ref{R}), the $u$- and $v$- components of the breather generated by the
symmetry-breaking instability are no longer mutually identical. This
manifestation of the symmetry-breaking instability is illustrated by Fig. %
\ref{fig5}(c), which displays the evolution of the difference between the
peak powers of the two components, and the separation between their centers.
The latter is defined as%
\begin{equation}
X_{u}-X_{v}\equiv \frac{1}{P_{u}}\int_{-\infty }^{+\infty }|u\left(
x,z\right) |^{2}xdx-\frac{1}{P_{v}}\int_{-\infty }^{+\infty }|v\left(
x,z\right) |^{2}xdx.  \label{X}
\end{equation}

It is relevant to mention that the second finite bandgap also contains a
branch of the so-called subfundamental solitons, whose power is smaller than
that of the fundamental GSs \cite{sub1,sub2,sub3}. These are odd modes,
squeezed, essentially, into a single cell of the underlying lattice
potential. The subfundamental \ solitons are unstable, tending to rearrange
themselves into fundamental ones belonging to the first finite bandgap,
therefore they are not considered below.
\begin{figure}[tbp]
\centering\includegraphics[width=3in]{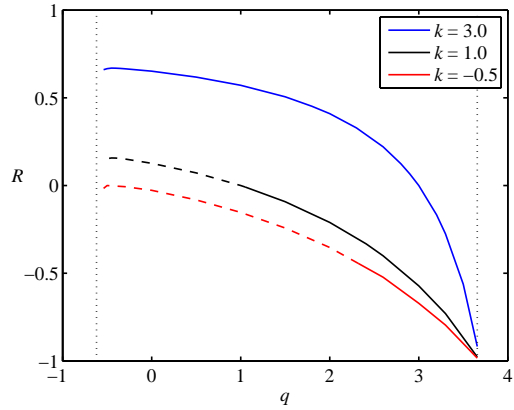}
\caption{The asymmetry ratio, $R$ [defined as per Eq. (\protect\ref{R})],
versus propagation constant $q$, at fixed values of $k=3.0$, $1.0$, and $%
k=-0.5$ (the top, middle, and bottom curves, respectively), for asymmetric
fundamental solitons of the intra-gap type. Stable and unstable branches are
shown by solid and dashed lines, respectively.}
\label{fig8}
\end{figure}
\begin{figure}[tbp]
\centering\includegraphics[width=2.0in]{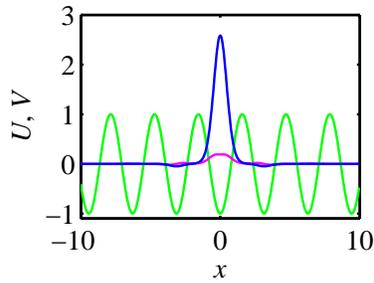}
\caption{An example of a stable strongly asymmetric soliton with $k=-0.5$
and $q=3.65$. Fields $U(x)$ and $V(x)$, which pertain to propagation
constants $k$ and $q$, are shown, respectively, by the magenta (lower) and
blue (taller) profiles.}
\label{fig9}
\end{figure}
\begin{figure}[tbph]
\centering\subfigure[]{\includegraphics[width=2.5in]{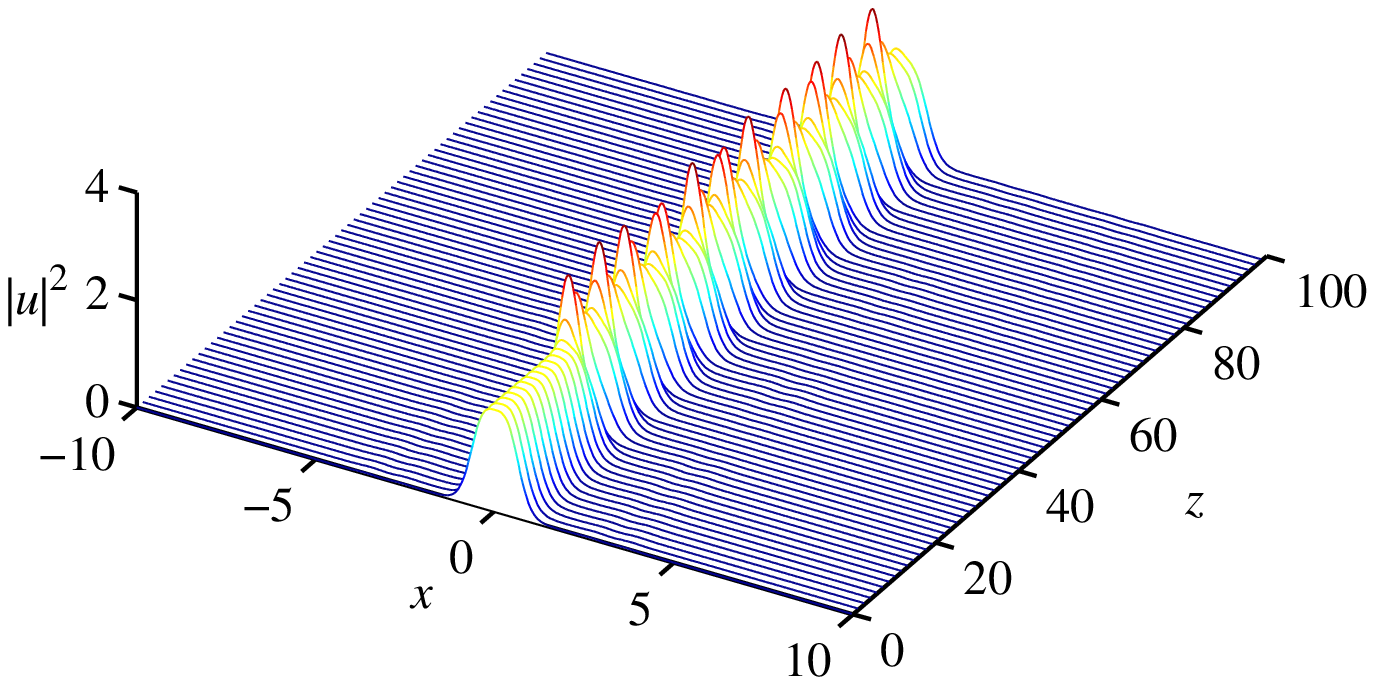}}%
\subfigure[]{\includegraphics[width=2.5in]{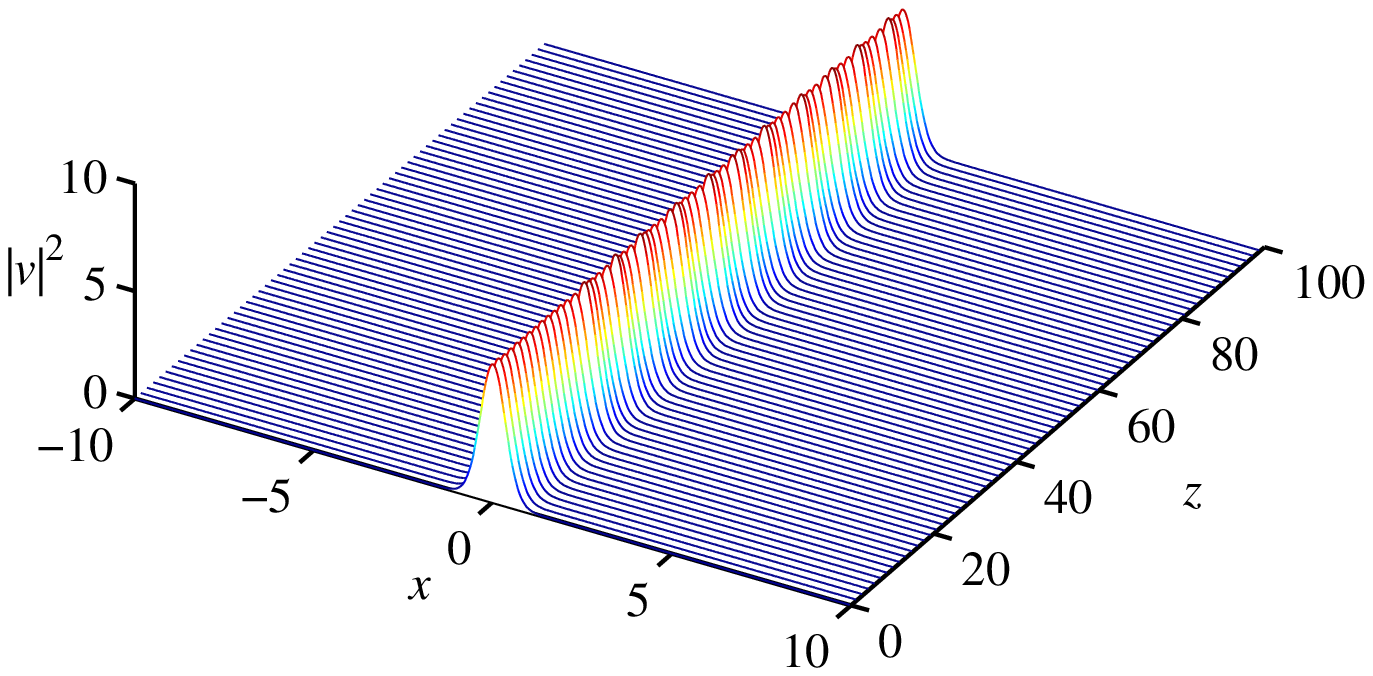}}
\caption{A typical example of the transformation of the unstable asymmetric
gap solitons into a breather, for $k=-0.5$ and $q=2.0$.}
\label{fig10}
\end{figure}

\subsection{Asymmetric solitons of the intra-gap type}

As said above, two-component asymmetric fundamental GSs, with different
propagation constants, $k\neq q$, may be naturally classified as solitons of
the intra- and inter-gap types if $k$ and $q$ belong to the same or
different finite bandgaps \cite{Arik}. In this subsection, we report results
for asymmetric intra-gap solitons with both $k$ and $q$ falling into the
first finite bandgap, as well as for asymmetric breathers developing from
such solitons when they are unstable.

Examples of stable asymmetric GSs of the intra-gap type are displayed in
Fig. \ref{fig7}, for a fixed asymmetry ratio, $R=-0.5$, defined as per Eq. (%
\ref{R}). The GS family, along with the family of persistent breathers into
which unstable solitons are spontaneously transformed, is represented in
Fig. \ref{fig8} by dependences $R(q)$ at different fixed values of the other
propagation constant, $k$.

It is possible to explain the fact that all the $R(q)$ curves converge to $%
R=-1$, as $q$ approaches the right edge of the bandgap in Fig. \ref{fig8}.
In this case, the $V$ component turns into the delocalized Bloch wave
function with a diverging power, $P_{v}$, that corresponds to $%
P_{u}/P_{v}\rightarrow 0$ [it is tantamount to $R\rightarrow -1,$ as per Eq.
(\ref{R})]. An example of a stable GS, close to this limit, with $k=-0.5$, $%
q=3.65$ and $R=-0.9811$, is shown in Fig. \ref{fig9}. The central core of
the $V$-component is described by the TFA, based on Eq. (\ref{in}), as the
corresponding necessary condition (\ref{<<}) holds in this case, while the
TFA does not apply to the $U$-component. The presence of undulating tails,
which are close to the Bloch functions, rather than the simple approximation
(\ref{outer}), which is valid far from the edge of the bandgap, is also
visible in Fig. \ref{fig8}.

Those asymmetric intra-gap GSs which form unstable subfamilies in Fig. \ref%
{fig8} are destabilized by oscillatory perturbations. The instability
transform the solitons into breathers, see a typical example in Fig. \ref%
{fig10} [cf. the examples of the destabilization of the symmetric GSs shown
in Fig. \ref{fig5}(b,c)]. The emerging breathers keep values of the
asymmetry ratio (\ref{R}) almost identical to those of their parent GSs; for
instance, in the case displayed in this figure, the unstable soliton with $%
R_{\mathrm{initial}}=-0.3617$\emph{\ }evolves into the breather with $R_{%
\mathrm{final}}=-0.3623$.

It is relevant to stress that the transformation of unstable stationary GSs
into the breathers gives rise to little radiation loss of the total power, $%
P $. On the other hand, in the general case a given unstable gap soliton
does not have a stable counterpart with a close value of $P$, hence this
unstable soliton cannot transform itself into a slightly excited state of
another stable GS. Thus, the breathers represent a distinct species of
localized modes.

\begin{figure}[tbp]
\centering\subfigure[]{\includegraphics[width=2.5in]{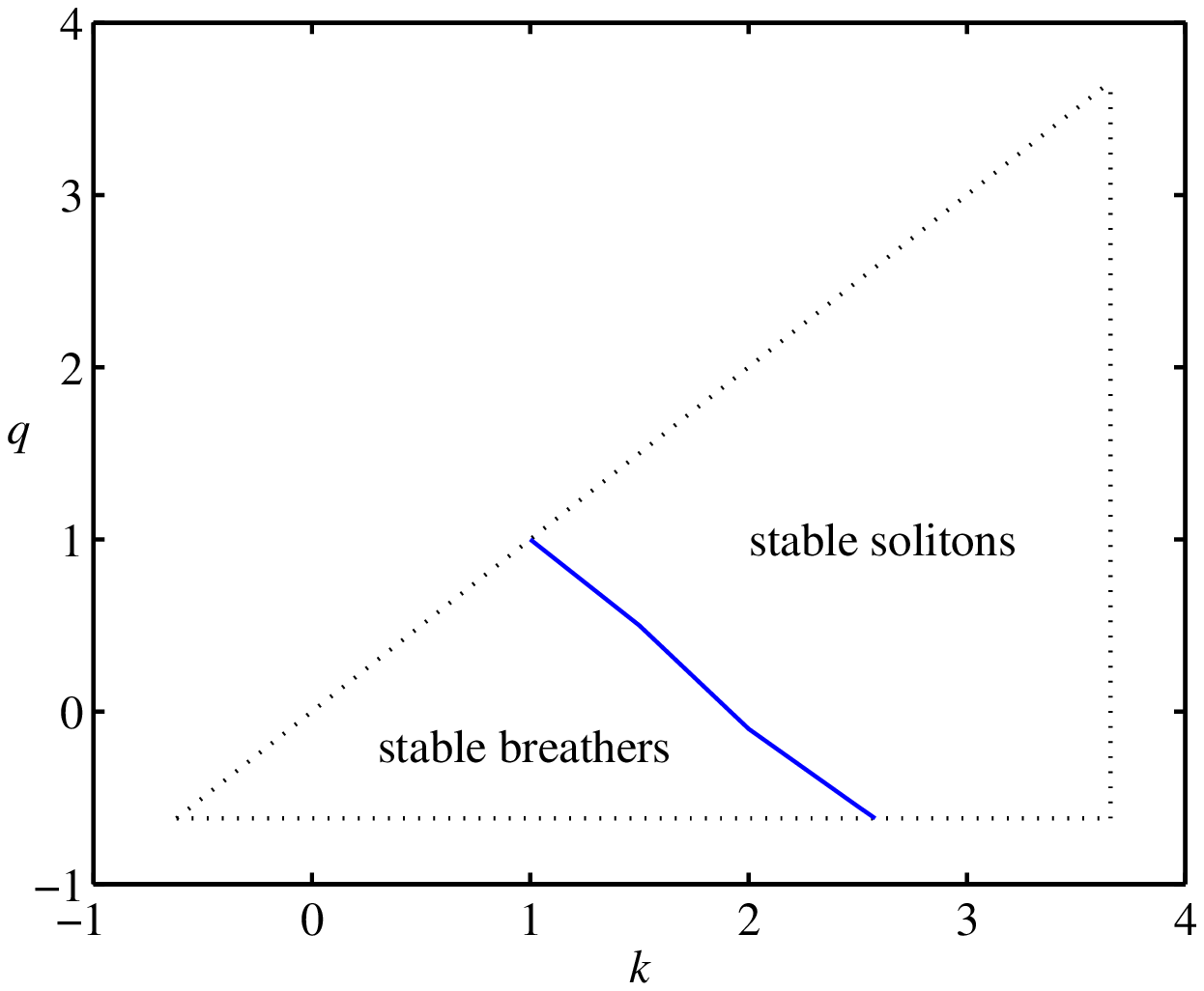}}%
\subfigure[]{\includegraphics[width=2.5in]{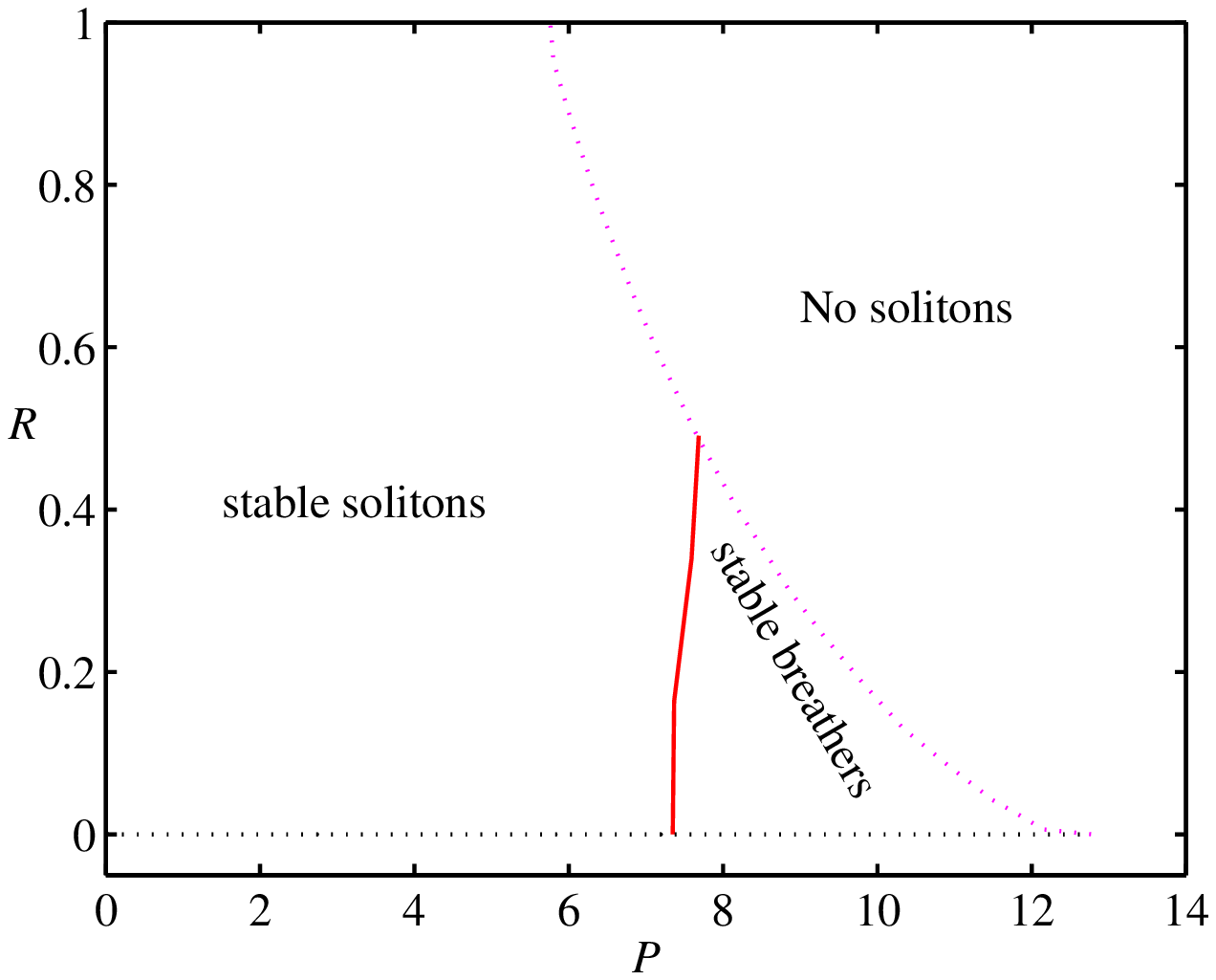}}
\caption{(a) The stability border in the plane of the propagation constants,
$\left( k,q\right) $, for asymmetric solitons of the intra-gap type. Only
half of the plane is shown, delineated by the dotted triangle, within which
wavenumbers $k$ and $q$ belong to the first finite bandgap, as the other
half is a mirror image of the displayed one. (b) The same in the plane of
the total power and asymmetry ratio, $\left( P,R\right) $, defined as per
Eqs. (\protect\ref{P}) and (\protect\ref{R}). Localized modes do not exist
above the right boundary of the stability regions in panel (b). The diagram
at $R<0$ is a mirror image of the one displayed here for $R>0$.}
\label{fig11}
\end{figure}

The most essential results of the stability analysis for the asymmetric
solitons of the intra-gap type, and for breathers replacing unstable
solitons, are summarized by diagrams in the planes of $\left( k,q\right) $
and $\left( P,R\right) $, which are displayed in Fig. \ref{fig11}. The
predictions of the analysis based on the computation of the stability
eigenvalues for the stationary solitons, as per Eqs. (\ref{u1})-(\ref{v2}),
always comply with stability tests provided by direct simulations of Eqs. (%
\ref{1Dpsi}) and (\ref{1Dphi}).

As mentioned above, the instability of a part of the branch of the symmetric
solitons along the line of $R=0$ in Fig. \ref{fig11}(b) implies that the
symmetry-breaking perturbations destabilize the symmetric solitons in the
first finite bandgap at $P>P_{\mathrm{cr}}$, see Eq. (\ref{cr}), while their
counterparts are stable in the single-component system. Another clear
conclusion is that the stability region gradually shrinks with the increase
of the asymmetry.

\subsection{Solitons of the inter-gap type}

\begin{figure}[tbp]
\centering\includegraphics[width=2in]{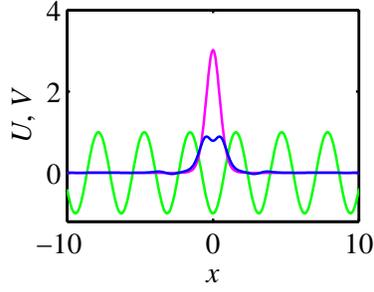}
\caption{An example of a stable inter-gap soliton, for $k=3$ and $q=-1.65$.
The single-peak and split-peak profiles, $U(x)$ and $V(x)$, represent,
respectively, the components in the first and second finite bandgaps.}
\label{fig12}
\end{figure}

\begin{figure}[tbp]
\centering\subfigure[]{\includegraphics[width=2.5in]{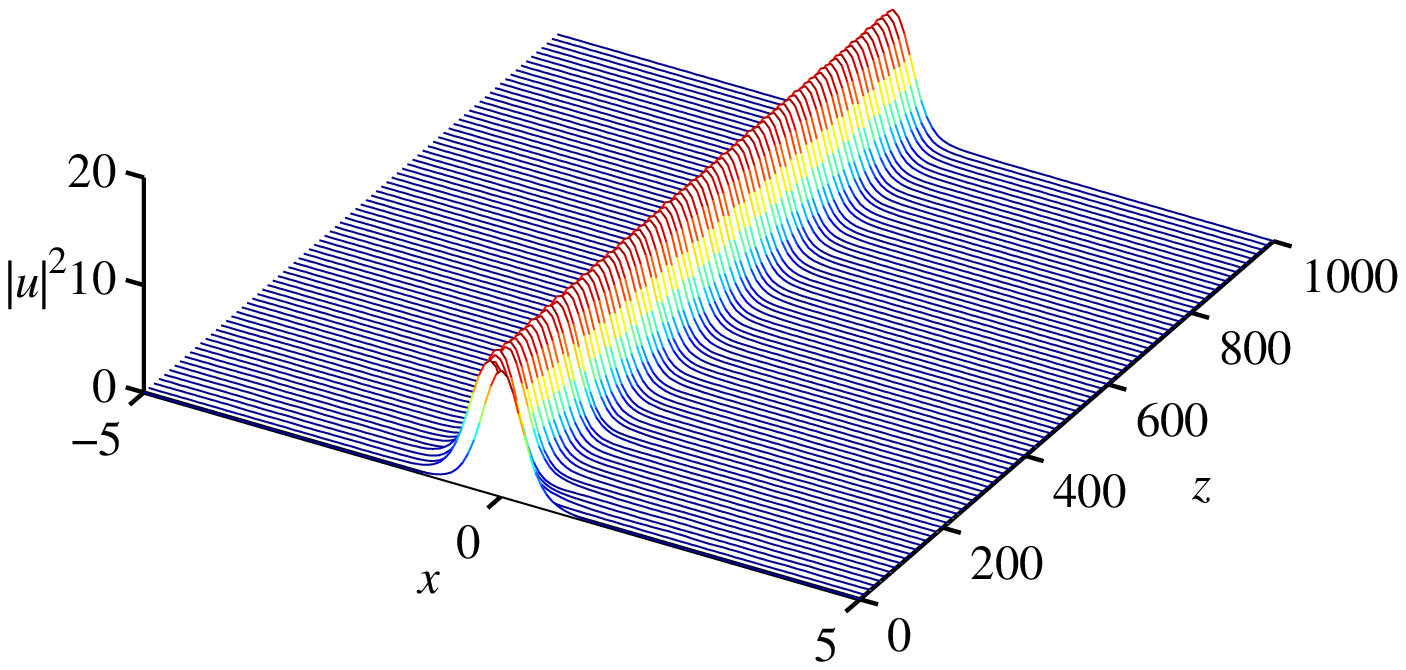}}%
\subfigure[]{\includegraphics[width=2.5in]{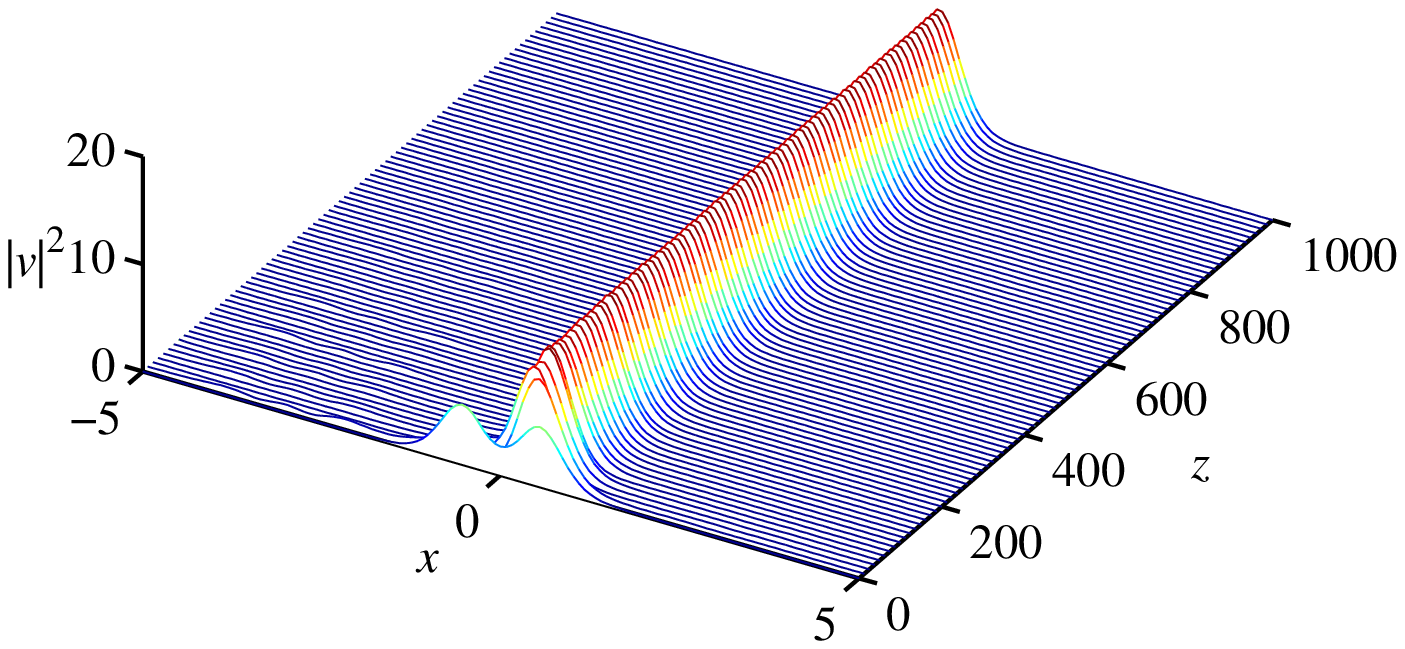}} \subfigure[]{%
\includegraphics[width=2.0in]{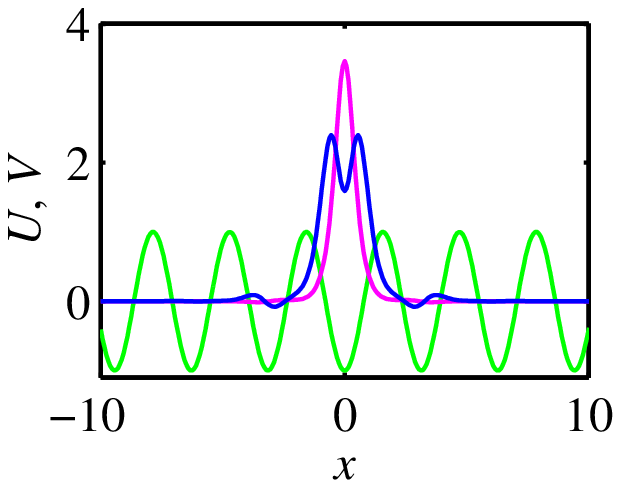}}
\caption{A typical example of the transformation of an unstable intergap
soliton into a stable breather, at $k=0$ and $q=-2$. (a,b) The evolution of $%
|u|^{2}$ and $|v|^{2}$. (c) The initial profiles of $U(x)$ and $V(x)$
(single-peak and split-peak shapes, respectively).}
\label{fig13}
\end{figure}

All the GSs of the inter-gap type, with two propagation constants belonging
to the two different finite bandgaps, are naturally asymmetric, even if
their components have equal powers. Examples of stable and unstable
inter-gap solitons are displayed in Figs. \ref{fig12} and \ref{fig13},
respectively. A noteworthy feature exhibited by these examples is a
split-peak structure of the component belonging to the second finite
bandgap.

\begin{figure}[tbp]
\centering\includegraphics[width=3in]{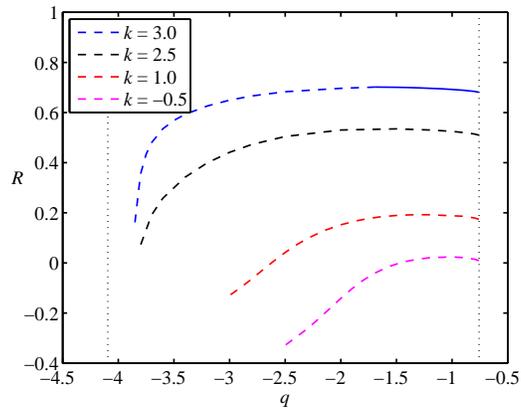}
\caption{Asymmetry ratio $R$ for the inter-gap solitons versus propagation
constant $q$ in the second finite bandgap, at fixed values $k=3.0$, $2.5$, $%
1.0$, and $-0.5$ (from the top to the bottom) of wavenumber $k$ in the first
bandgap. Solid and dashed lines designate stable stationary solitons and
breathers, respectively.}
\label{fig14}
\end{figure}

The asymmetry measure, $R(q)$, for families of the inter-gap solitons is
plotted in Fig. \ref{fig14}(a) versus the propagation constant $q$ in the
second bandgap, at fixed values of $k$ (the propagation constant in the
first bandgap). The stability of the respective GS families is also shown in
Fig. \ref{fig14}.

\begin{figure}[tbp]
\centering\subfigure[]{\includegraphics[width=2.5in]{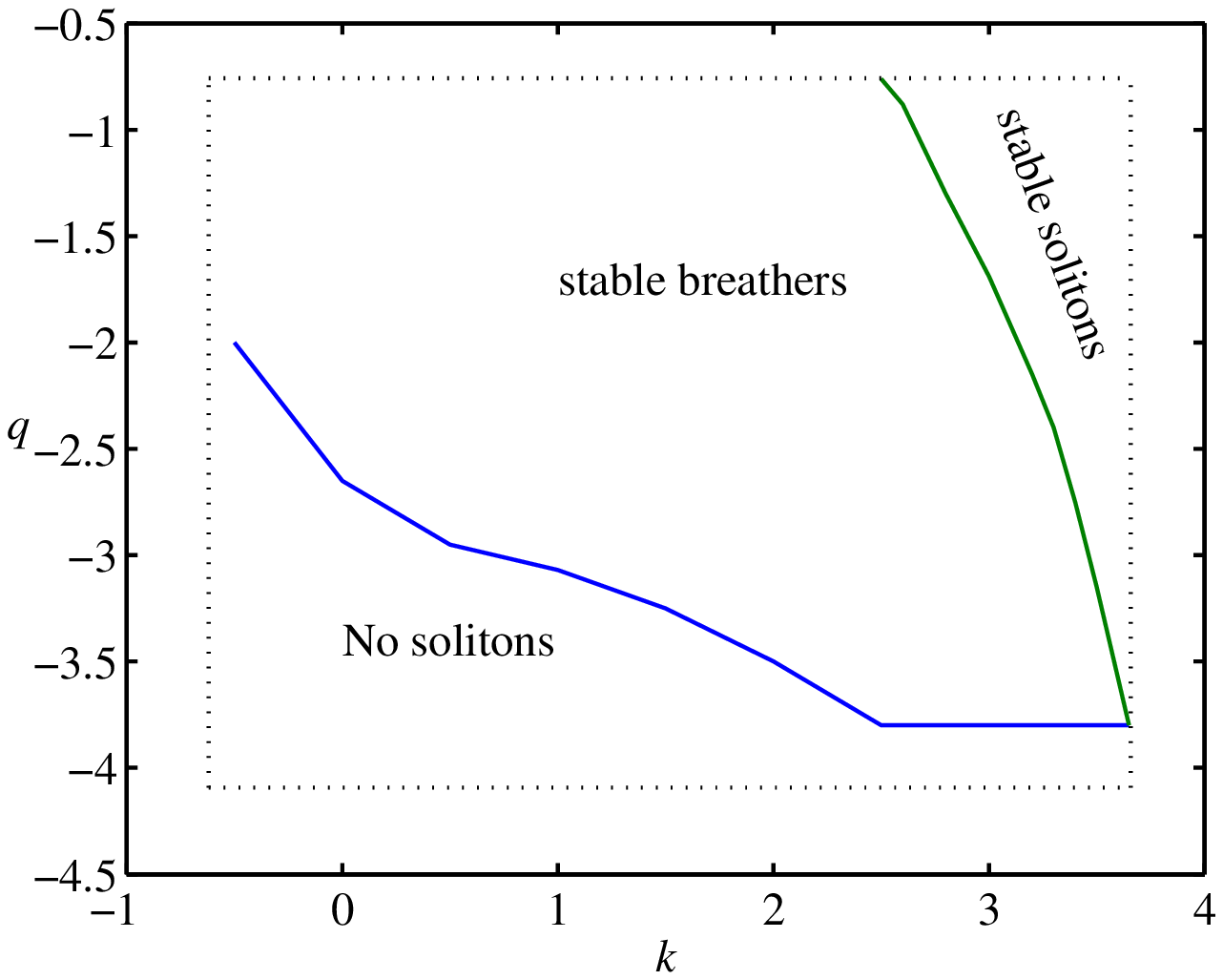}}%
\subfigure[]{\includegraphics[width=2.5in]{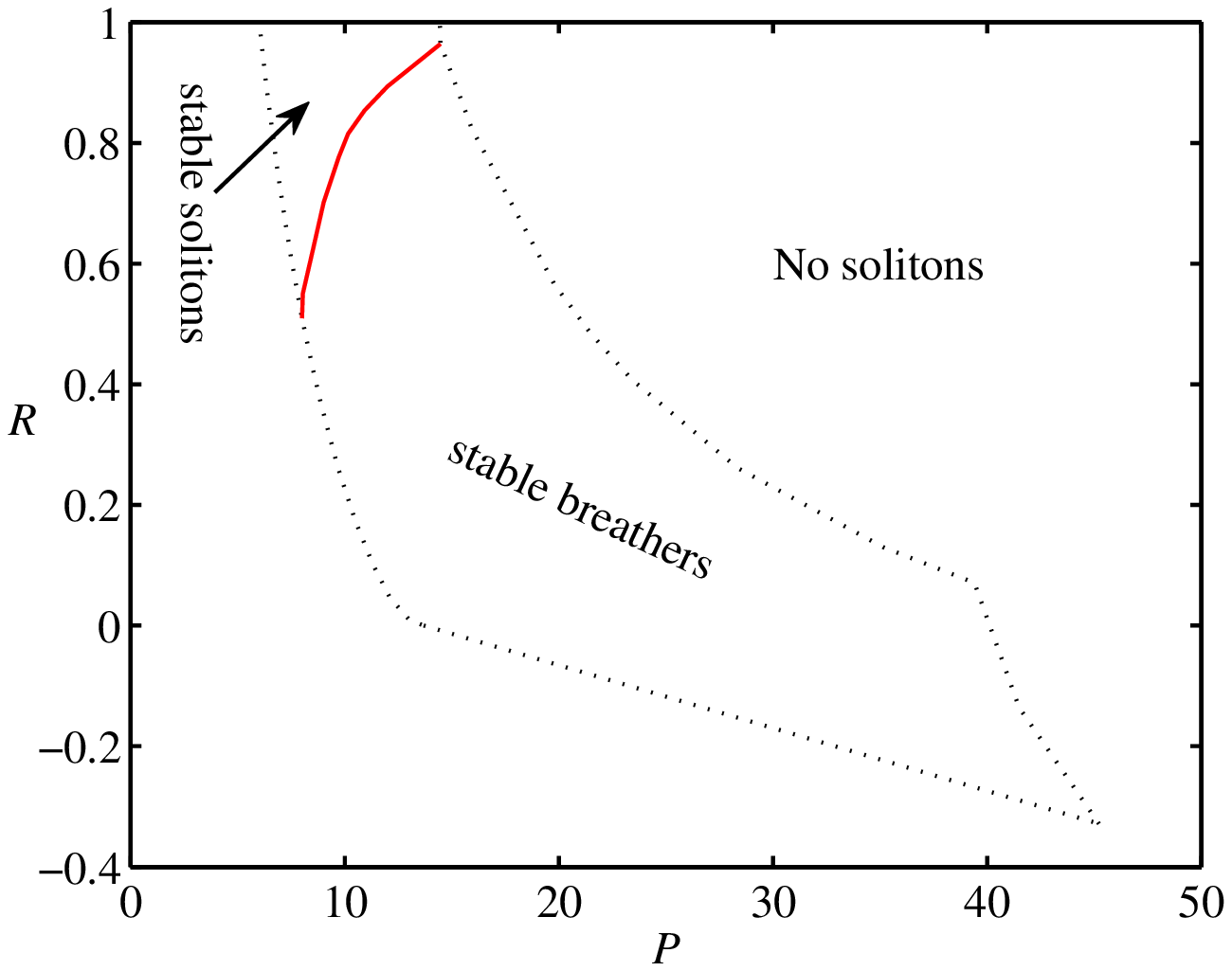}}
\caption{The same as in Fig. \protect\ref{fig11}, but for inter-gap
solitons. In (a), the dotted rectangle delineates the region occupied by
wavenumbers $k$ and $q$ belonging to the first and second finite bandgaps,
respectively.}
\label{fig15}
\end{figure}

The (in)stability of the inter-gap solitons is summarized by the diagrams in
the planes of $\left( k,q\right) $ and $\left( P,R\right) $ presented in
Fig. \ref{fig15}, cf. similar diagrams for intra-gap solitons shown above in
Fig. \ref{fig11}. As well as in that case, unstable inter-gap solitons are
spontaneously replaced by robust localized breathers. The spontaneous
transformation increases the initial degree of the asymmetry:\ For instance,
an unstable inter-gap soliton with $R=0.019$ is converted into a breather
with $R=0.028$.

In the present case too, the existence region of stable modes shrinks with
the increase of the asymmetry; note also that the stationary inter-gap
solitons may be stable solely at sufficiently large values of the asymmetry,
$R\geq R_{\min }\approx 0.5$. The asymmetric shape of the stability diagram
in Fig. \ref{fig15}(b) with respect to $R>0$ and $R<0$ [unlike the symmetry
of the diagram for the intra-gap solitons implied in Fig. \ref{fig11}(b)] is
explained by the fact that, in definition (\ref{R}), $R>0$ implies that the
dominant component resides in the first finite bandgap, where it is more
robust than in the second bandgap.

Finally, out additional analysis has demonstrated that all the stationary
GSs---not only the symmetric ones [see Fig. \ref{fig2}], but also all the
asymmetric solitons of the intra-gap type---are completely unstable in the
second finite bandgap. They too tend to spontaneously rearrange themselves
into breathers, which is not shown here in detail.

\section{Conclusion}

We have introduced the model of symbiotic two-component GSs (gap solitons),
based on two nonlinear Schr\"{o}dinger equations coupled by the repulsive
XPM terms and including the lattice potential acting on both components, in
the absence of the SPM nonlinearity. The model has a realization in optics,
in terms of \textquotedblleft holographic solitons" in photonic crystals,
and as a model of binary quantum gases (in particular, a fully polarized
fermionic one) loaded into the optical-lattice potential. Families of
fundamental asymmetric GSs have been constructed in the two lowest finite
bandgaps, including the modes of both the intra-gap and inter-gap types,
i.e., those with the propagation constants of the two components belonging
to the same or different bandgaps, respectively. The existence and stability
regions of the symbiotic GSs and breathers, into which unstable solitons are
transformed, have been identified. A noteworthy finding is that
symmetry-breaking perturbations destabilize the symmetric GSs in the first
finite bandgap, if their total power exceeds the critical value given by Eq.
(\ref{cr}), along with all the symmetric solitons in the second bandgaps. It
was demonstrated too that the stability area for the intra-gap GSs shrinks
with the increase of the asymmetry ratio, $R$. On the other hand, inter-gap
GSs may be stable only for sufficiently large ratio, $R>0.5$. The intra-gap
solitons are completely unstable in the second bandgap. Some features of the
GS families were explained by means of the extended TFA (Thomas-Fermi
approximation), augmented by the tails attached to the taller component, in
the case of asymmetric solitons.

A natural extension of the analysis may deal with 2D symbiotic gap solitons,
supported by the square- or radial-lattice potentials. In that case, it may
be interesting to consider two-component solitary vortices too.

\section*{Acknowledgment}

The work of T.M. was supported by the Thailand Research Fund through grant
RMU5380005. B.A.M. appreciates hospitality of the Mahanakorn University of
Technology (Bangkok, Thailand).

\end{document}